\documentclass[11pt]{article}
\usepackage{hyperref}
\pdfoutput=1
\begin{document}
\title{Bow wave and spray dynamics by a wedge}
\author{Zhaoyuan Wang, Jianming Yang, Frederick Stern\\
\\\vspace{6pt} IIHR-Hydroscience \& Engineering, \\ University of Iowa, Iowa city, IA 52242, USA}
\maketitle
\begin{abstract}
Flows around a wedge-shaped bow are simulated with the aim of investigating the wave breaking mechanism and small scale features of ship bow waves. This fluid dynamics video shows the plunging wave breaking process around the wedge including the thin water sheet formation, overturning sheet with surface disturbance, fingering and breaking up into spray, plunging and splashing, and air entrainment. 
\end{abstract}
\section{Introduction}
Ship bow waves exhibit both large and small scale features. The most prominent small scale feature is the bow wave crest formation of thin overturning sheets which break up into spray. Re and We scale effects are large such that replication of full scale phenomena of the small scale features of ship bow waves is difficult even with large models. However, experimental studies for wedge flows display and document the small scale structures of bow waves.  Herein, flows around a wedge-shaped bow are numerically simulated with the aim of investigating the wave breaking mechanism and small scale features of ship bow waves.
The side length of the wedge is L = 0.75 m, and the height of the wedge is H = 1.0 m. The half wedge angle is $\theta$ = $26^\circ$ and the flare angle $\phi$ = $0^\circ$. The water depth is d = 0.0745 m and the upstream velocity is U = 2.5 m/s, the corresponding Reynolds number,  Re = $1.64\times{10}^5$, and the Froude number, Fr = 2.93. 

\end{document}